\documentclass [a4paper, 12pt] {article}
\usepackage [english] {babel}
\usepackage [dvips] {graphicx}
\usepackage {longtable}

\font\msbm=msbm10 scaled \magstep1 

\def\Bbb*1 {\mbox {\msbm *1} \,}
\def\beq {\begin {equation}}
\def\eeq {\end {equation}}
\def\bdm {\begin {displaymath}}
\def\edm {\end {displaymath}}
\def\bea {\begin {eqnarray}}
\def\eea {\end {eqnarray}}
\def\ba {\begin {array} {l}}
\def\ea {\end {array}}

\begin {document}
\title {EXPLICIT SOLUTIONS TO BOUNDARY PROBLEMS FOR 2+1-DIMENSIONAL INTEGRABLE SYSTEMS.}
\author {V.L. Vereschagin \\
Institute of Mathematics Ufa Sci. Centre RAS}
\date {\null}
\frenchspacing \maketitle\

\begin {abstract}
Nonlinear integrable models with two spatial and one temporal variables: Kadomtsev-Petviashvili equation and two-dimensional Toda lattice are investigated on the subject of correct formulation for boundary problem that can be solved within the framework of the Inverse Scattering Problem method. It is shown that there exists a large set of integrable boundary problems and various curves can be chosen as boundary contours for them. We develop a method for obtaining explicit solutions to integrable boundary problems and its effectiveness is illustrated by series of examples.
\end {abstract}

\section {Introduction}.

Known examples of integrable 2+1-dimensional\footnote{Usually dependence on two spatial and one temporal variables is implied} systems are still not numerous, but they attract considerable interest in the aspect of the theory of differential equations as well as of various applications. Probably the most noteworthy example is Kadomtsev-Petviashvili equation (KP, see \cite{kadpet})

\begin{equation}
\label{1.1}
\begin{array}{c}
u_T+u_{xxx}-6uu_x=-3\alpha ^2 w_Y\\
w_x=u_Y,
\end{array} 
\end{equation}
where $u=u(x,Y,T),\ w=w(x,Y,T),\ u_x=\frac{\partial u}{\partial x},\ x,\ Y,\ T$ are continuous variables, parameter $\alpha$ equals either $1$ or $i$.

Another important example is presented by difference-differential model - two-dimensional (2D) Toda lattice (see \cite{mik1})

\begin{equation}
\label{1.2}
 v_{XY}(X,Y,n)=W(X,Y,n-1)-W(X,Y,n), 
\end{equation}
where $v(X,Y,n)$ is an unknown function of two continuous variables $X,\ Y$ and integer index $n\in {\bf Z}$, $W(X,Y,n)=exp (v(X,Y,n)-v(X,Y,n+1))$.

The phenomenon of integrability of such systems is based on existence of so-called commutator representations or Lax representations. Examine this concept on the example of KP system (\ref {1.1}). It was shown (see papers \cite{teorsol}, \cite{dr}, \cite{zash}) that nonlinear equation (\ref {1.1}) constitutes consistency condition for two linear equations of the following type:

\begin{equation}
\label{1.3}
\begin{array}{c}
\alpha \Psi_Y=\Psi_{xx}-u\Psi=L\Psi\\
\Psi_T=-4\Psi_{xxx}+6u\Psi_x+3(u_x+\alpha w)\Psi=A\Psi,
\end{array} 
\end{equation}
which specify differential Lax operators $L,\ A$ so that KP equation is written in the commutator form

\begin{equation}
\label{1.4}
[\partial _Y-L,\partial _T-A]=0, 
\end{equation}
where $\partial _Y=\partial /\partial Y$, $[,]$ is commutator. For this reason KP equation (\ref {1.1}) may be referred to as integrable model which possess a large class of solutions important for both theory and applications. There exist soliton solutions to KP that can be obtained with the help of the ``dressing'' method (see \cite{teorsol}, \cite{zash}).

The special interest from the point of view of applications is inspired by possibility to solve boundary problems for 2+1-dimensional models. It must be mentioned that not any boundary problem for KP is solvable in the sense of presence of large enough set of representative ``nice'' solutions like those solitons. The point is that by imposing an arbitrary boundary constraint one brakes the standard scheme of the Inverse Scattering Method (ISM) since the infinite hierarchy of integrals of motion vanishes. So we arrive at the question if there exist formulations of boundary problem which save sufficiently many integrals of motion to keep up applicability of the ISM. Considerable progress in this aspect was provided by papers \cite{skl}, \cite{hab1}, \cite{hab2}, \cite{hab3}, \cite{aggh}. The following Definition constitutes the condition which allows to select boundary problems compatible with the ISM.

\underline{Definition 1.} Let nonlinear equation admit two different Lax representations not connected via any gauge transformation:

$$
\{\Psi_X=A\Psi,\ \Psi_Y=L\Psi \},\ \{\widehat \Psi_X=\widehat A\widehat \Psi,\ \widehat \Psi_Y=\widehat L\widehat \Psi \}.
$$
Then the boundary constraint $\Omega (u)=0$ on the boundary $X=X_0$ is called integrable if there exists a linear differential operator $B$ so that on boundary $X=X_0$ function $\widehat \Psi=B\Psi$ solves the equation $\widehat \Psi_Y=\widehat L\widehat \Psi$ for any solution $\Psi$ of equation $\Psi_Y=L\Psi$ as the constraint $\Omega (u)=0$ holds.

In papers \cite{hab2}, \cite{hab3} the Definition 1 was, in partial, used to answer the question of correct formulation of boundary problems for KP equation on straight boundary contour $(X=X_0)$, similar questions for 2D Toda lattice with circular boundary contour were investigated in paper \cite{ghz}. Nevertheless it was unclear how to determine the class of integrable (in the sense of Definition 1) boundary problems, which curves may be chosen as appropriate boundary contours.

This paper's goal is to develop a method that constructively determines integrable boundary problems for 2+1-dimensional models, KP equation (\ref {1.1}) and 2D Toda lattice (\ref {1.2}) form main examples. We present an algorithm for integration of corresponding boundary problems and the procedure yields solutions in the explicit form.

In section 2 we show main integrability attributes for equation (\ref {1.1}): Gelfand-Levitan-Marchenko (GLM) system associated with Lax representation (\ref {1.3}) and dressing method. In section 3 we rewrite equation (\ref {1.1}) in other variables and determine appropriate boundary contour and boundary constraint in generic form. In section 4 the mentioned tools are used to solve concrete boundary problem: we write out the GLM equation and find its kernel that enjoys the integrability condition. Then we utilize the dressing method to find solutions to boundary problem in the explicit form. The last part of the paper applies the developed method to solving boundary problems for 2D Toda lattice (\ref {1.2}).

\section{The use of the ISM.}
\setcounter{equation}{0}

Remind, following \cite{teorsol}, \cite{zash}, Zakharov-Shabat dressing method, one of ways to obtain explicit formulae for solutions to KP equation. Introduce operator $\widehat F$ such that

\begin{equation}
\label{2.1}
(\widehat F \Psi)(x)= \int ^\infty _{- \infty} F(x,z,T,Y)\Psi (z)dz,
\end{equation}

\begin{equation}
\label{2.2}
[\partial _T - A,\ F]=0,\ \ [\partial _Y - L,\ F]=0,
\end{equation}
and suppose that it admits the triangular factorization, i.e. the following form:

\begin{equation}
\label{2.3}
1+\widehat F=(1+\widehat K^+)^{-1}(1+\widehat K),
\end{equation}
where

\begin{equation}
\label{2.4}
\widehat K \Psi(x)= \int _{-\infty}^x K(x,z,T,Y)\Psi (z)dz,\ \ 
\widehat K^+ \Psi(x)= \int ^{\infty}_x K^+(x,z,T,Y)\Psi (z)dz.
 \end{equation}
Now examine operators $L^1$ è $A^1$:

\begin{equation}
\label{2.41}
\begin{array}{c}
L^1=\partial _x^2 -u^1 \\
A^1=-4\partial _x^3+u^1\partial _x+3(u^1_x+\alpha w^1_x),
\end{array}
\end{equation}
specified by the following constraints:

\begin{equation}
\label{2.5}
\begin{array}{c}
(\partial _T -A^1)(1+K)= (1+K) (\partial _T -A), \\
(\partial _Y -L^1)(1+K)= (1+K) (\partial _Y -L).
\end{array}
\end{equation}
They satisfy the commutator relation (\ref {1.4}), have the structure similar to that of operators $L,\ A$ and their coefficients can be found with the help of equations (\ref {2.5}), so that functions $u^1$ form new solutions to KP equation via the following formula:

\begin{equation}
\label{2.6}
u^1=u+2\partial _x K(x,x,Y,T)
\end{equation}
To find the function $K(x,z)$\footnote{For the sake of brevity we omit the dependence on variables $Y,\ T$.} we need to solve GLM equation which arises from systems (\ref {2.3}) and (\ref {2.4}):

\begin{equation}
\label{2.7}
K(x,z)+F(x,z)+\int ^{x}_{-\infty} K(x,z')F(z',z)dz'=0.
\end{equation}
Thus the dressing procedure consists of the following: we start from a simple potential $u(x)$ and solve linear equations (\ref {2.2}) on kernel $F(x,z)$. Then substitute $F(x,z)$ into GLM equation (\ref {2.7}), find $ K(x,z)$ which via formula (\ref {2.6}) yields new ``dressed'' solution to equation (\ref {1.1}).

Rewrite system (\ref {2.2}) in the following form:

\begin{equation}
\label{2.8}
\begin{array}{c}
\alpha F_{Y}-F_{xx}+F_{zz}+(u(x)-u(z))F=0\\
F_{T}+4F_{xxx}+4F_{zzz}-6u(x)F_x-6u(z)F_z\\ 
+3[\alpha (w(z)-w(x))-u_x(x)-u_z(z)]F=0.
\end{array}
\end{equation}
Look for solution to system (\ref {2.8}) in the form of product $F(x,z)=\Psi (x)\widehat \Psi(z)$. It is easy to verify that function $\Psi (x)$ solves system (\ref {1.3}) whereas $\widehat \Psi (x)$ solves system

\begin{equation}
\label{2.8a}
\begin{array}{c}
-\alpha \widehat\Psi _Y= \widehat\Psi _{xx}-u\widehat\Psi \\
\widehat\Psi _T= -4 \widehat\Psi _{xxx}+6u \widehat\Psi _{x} +3(u_x-\alpha w)\widehat\Psi ,
\end{array}
\end{equation}
which we call conjugate to system (\ref {1.3}). Then substitute just obtained kernel $F (x,z)$ into GLM equation (\ref {2.7}) and deduce formula for $K (x,z)$:

\begin{equation}
\label{2.9}
K(x,z)=-\frac{\Psi (x) \widehat \Psi (z)}{1+\int _{-\infty}^{x} \Psi (z')\widehat \Psi (z')dz'},
\end{equation}
which with the help of (\ref {2.6}) yields the ``dressed'' solution to KP.

The separate question is how to select solutions to integrable boundary problems among all the dressed solutions we obtained above. This will be done further via Definition 1.

\section{Integrable boundary problems.}
\setcounter{equation}{0}

Having the operator $L-A$-pair (\ref {1.3}) one can rewrite it via variable transform $Y=Y(y,t),\ T=T(y,t)$ so that integrable through Definition 1 boundary constraint is posed on straight contour $y=y_0$ or, in former variables, on curve line

\begin{equation}
\label{3.1}
Y=Y(y_0,t),\ T=T(y_0,t).
\end{equation}
In this case the system of linear equations (\ref {1.3}) looks in the following way:

\begin{equation}
\label{3.2}
\begin{array}{c}
\Psi _y = \frac {1}{\Delta}\left\{ \frac{f_{22}}{\alpha}(\Psi _{xx}-u\Psi)- f_{12}(-4 \Psi _{xxx}+6u \Psi _{x}+3(u_x+\alpha w))\Psi\right \} \\
\Psi _t = -\frac {1}{\Delta}\left\{ \frac{f_{21}}{\alpha}(\Psi _{xx}-u\Psi)- f_{11}(-4 \Psi _{xxx}+6u \Psi _{x}+3(u_x+\alpha w))\Psi \right \},
\end{array}
\end{equation}
where $f_{ij}$ are partial derivatives: 

$$
f_{11}=y_Y=\frac{T_t}{T_tY_y-Y_tT_y},\ f_{12}=t_Y=\frac{T_y}{T_yY_t-Y_yT_t},
$$
$$
f_{21}=y_T=\frac{Y_t}{T_yY_t-T_tY_y},\ f_{22}=t_T=\frac{Y_y}{T_tY_y-T_yY_t},\ \Delta = f_{11}f_{22}-f_{12}f_{21},
$$
and conjugate system (\ref {2.8a}) in new variables is obtained from (\ref {3.2}) by simple replacement $\alpha \rightarrow -\alpha$:
\begin{equation}
\label{3.3}
\begin{array}{c}
\widehat\Psi _y = -\frac {1}{\Delta}\left\{ \frac{f_{22}}{\alpha}(\widehat\Psi _{xx}-u\widehat \Psi)+ f_{12}(-4 \widehat\Psi _{xxx}+6u \widehat \Psi _{x}+3(u_x-\alpha w))\widehat \Psi\right \} \\
\widehat\Psi _t = \frac {1}{\Delta}\left\{ \frac{f_{21}}{\alpha}(\widehat\Psi _{xx}-u\widehat\Psi)+ f_{11}(-4 \widehat\Psi _{xxx}+6u \widehat \Psi _{x}+3(u_x-\alpha w))\widehat \Psi \right \}.
\end{array}
\end{equation}
We associate the system (\ref {3.2}) and its conjugate one (\ref {3.3}) with the same solution $u(x)$ of integrable boundary problem, so in virtue of Definition 1 we demand that linear correlation $\widehat\Psi (x)=B(x,t)\Psi (x)\vert _{y=y_0}$\footnote{In general $B$ is a differential operator, here for simplicity we restrict to the case of $B$ being a constant multiplier.} holds on the boundary contour. Substitute this into the second equation of system (\ref {3.3}), with the help of (\ref {3.2}) collect coefficients at functions $\Psi,\ \Psi _{x},\ \Psi _{xx},\ \Psi _{xxx}$ and arrive at explicit formulae for the multiplier $B(x,t)$ :

\begin{equation}
\label{3.4}
B(x,t)=g(t)e^{\frac{xf_{21}}{6\alpha f_{11}}} ,
\end{equation}
where $g(t)$ is arbitrary function, and for the boundary constraint:

\begin{equation}
\label{3.5}
\left. \Delta (\log g(t))_t + \frac{x\Delta}{6\alpha}\left(\frac{f_{21}}{f_{11}}\right)_t+\frac{f_{21}u}{\alpha} +6\alpha f_{11}w-\frac{f_{21}^3}{108 \alpha ^3 f_{11}^2} \right |_{y=y_0}=0.
\end{equation}
We must also demand that the following conditions on the variable transformation hold:

\begin{equation}
\label{3.6}
f_{11}\ne 0,\ f_{21}\ne 0,\ \Delta \ne 0.
\end{equation}

\section{Explicit formulae for solutions to integrable boundary problems.}
\setcounter{equation}{0}

The variable transformations $Y=Y(y,t),\ T=T(y,t)$ specify boundary contours as parametric curves (\ref {3.1}). For example we choose the following simple transformation:

\begin{equation}
\label{4.1}
Y=yt,\ \ T=\frac{y}{t},
\end{equation}
which determines the boundary contour as hyperbolic curve $YT=y_0^2$. In this case the integrable boundary constraint (\ref {3.5}) is presented by formula
\begin{equation}
\label{4.2}
\left. -\frac{x}{3T}+u+6\alpha ^2 \frac{T}{Y}w-\frac{\alpha ^2}{108}\frac{Y^2}{T^2}\right | _{YT=y^2_0}=0,
\end{equation}
for simplicity $g(t)\equiv const$.

Let the following elementary solution of the boundary problem (\ref {1.1}), (\ref {4.2}) be the initial step for the dressing procedure:
 
\begin{equation}
\label{4.3}
u^0=\frac{\alpha ^2 Yy^2_0}{18T^3},\ w^0 =\frac{\alpha ^2 xy^2_0}{18T^3}+\frac{Y^2y^2_0}{36T^4}-\frac{23y^6_0}{648T^6}
\end{equation}
Substitute solution (\ref {4.3}) into coefficients of linear systems (\ref {3.2}), (\ref {3.3}) and find solutions of these systems:
 
\begin{equation}
\label{4.4}
\begin{array}{c}
\Psi = \sum ^N _{j=1} d_j\exp \left [ x\left (p_j -\frac{y_0^2}{12\alpha T^2}\right )+\frac{Y}{\alpha} \left (p_j -\frac{y_0^2}{12\alpha T^2}\right )^2 \right. \\
\left. -\frac{\alpha Y^2 y_0^2}{36 T^3}+\frac{\alpha y_0^6}{48 T^5}+ \frac{\alpha ^2 y_0^4 p_j}{36 T^3}-\frac{ y_0^2 p_j^2}{\alpha T}-4Tp_j^2 \right ]
\end{array}
\end{equation}

\begin{equation}
\label{4.5}
\begin{array}{c}
\widehat \Psi = \sum ^N _{j=1} l_j\exp \left [ x\left (q_j +\frac{y_0^2}{12\alpha T^2}\right )-\frac{Y}{\alpha} \left (q_j +\frac{y_0^2}{12\alpha T^2}\right )^2 \right. \\
\left. +\frac{\alpha Y^2 y_0^2}{36 T^3}-\frac{\alpha y_0^6}{48 T^5}+ \frac{\alpha ^2 y_0^4 q_j}{36 T^3}+\frac{ y_0^2 q_j^2}{\alpha T}-4Tq_j^3 \right ],
\end{array}
\end{equation}
where $d_j,\ l_j,\ p_j,\ q_j,\ j=1,2,...,N$ are constant parameters. Choose the simplest case $N=1$ which may be formally associated with one-soliton solution. To keep the integrability one must satisfy Definition 1, i.e. the following linear condition on the boundary:
 
\begin{equation}
\label{4.6}
\widehat \Psi = B \Psi \vert _{y=y_0},\ \ B=e^{\frac{xt^2}{6\alpha}}.
\end{equation}
So we must take solutions (\ref {4.4}), (\ref {4.5}) of linear systems (\ref {3.2}), (\ref {3.3}) and select those solutions which satisfy the condition (\ref {4.6}). For $N=1$ and $p=q$ this condition holds. Then substitute the appropriate functions $\Psi$ and $\widehat \Psi$ into the formula for solutions to GLM equation:

\begin{equation}
\label{4.7}
K(x,x)=-\frac{\Psi (x)\widehat \Psi (x)}{1+\int _{-\infty} ^{x}\Psi (z)\widehat \Psi (z)dz}=-\left(\frac{1}{2p}+\chi \right)^{-1},
\end{equation}
where
 
\begin{equation}
\label{4.8}
\chi =\exp \left(-2xp +\frac{\alpha ^2Yy_0^2p}{3T^2}-\frac{\alpha ^2y_0^4p}{18T^3}+8Tp^3\right),
\end{equation}
and thus in force of formula(\ref {2.6}) obtain the ``dressed'' solution to KP:
 
\begin{equation}
\label{4.9}
\begin{array}{c}
u^1= \frac{\alpha ^2Yy_0^2}{18T^3}-\frac{2p^2}{\cosh ^2\tau},\\
w^1= \frac{\alpha ^2y_0^2x}{18T^3}+\frac{\alpha ^2y_0^2p^2}{3T^2\cosh ^2\tau}+\frac{Y^2y_0^2}{36T^4}-\frac{23 y_0^6}{648T^6},
\end{array}
\end{equation}
where

$$
\tau =-px+ \frac{\alpha ^2Yy_0^2p}{4T^2}-\frac{\alpha ^2y_0^4p}{36T^3}+4Tp^3+\frac{1}{2}\log (2p).
$$
Note that putting $y_0=0$ we obtain straight boundary contour $Y=0$, linear phase $\tau$ and the solution (\ref {4.9}) turns into an ordinary soliton solution of KP and Korteweg-de Vries equations.
 \section{Boundary problems for discrete problems.}
\setcounter{equation}{0}

The method we developed above can be successfully applied to solving boundary problems for integrable 2+1-dimensional chains. Examine this procedure on the example of 2D Toda lattice (\ref {1.2}). For convenience rewrite it in other notations:

\begin{equation}
\label{1.2a}
 u_{XY}(X,Y,n)=w(X,Y,n-1)-w(X,Y,n), 
\end{equation}
where $w(X,Y,n)=exp (u(X,Y,n)-u(X,Y,n+1))$.

This model was proposed in paper \cite{mik1} together with appropriate Lax representation and integrals of motion. The nonlinear chain (\ref {1.2a}) presents the consistency condition for the following pair of linear equations:

\begin{equation}
\label{5.0}
\begin{array}{c}
\Psi _X (n)= -u_X (n) \Psi (n)+\Psi(n+1)=A^0 \Psi (n),\\
\Psi _Y (n)= -w(n-1) \Psi (n-1)=L^0 \Psi (n),
\end{array}
\end{equation}
where $L^0,\ A^0$ are difference Lax operators. Authors of papers \cite{mik2}, \cite{ueta} used this fact to develop approach to systems like (\ref {1.2a}) from the Inverse Scattering Problem viewpoint. Papers \cite{naka}, \cite{bms} presented the dressing method for 2D Toda lattice. This allowed to obtain a large class of explicit solutions to equation (\ref {1.2a}), the well-known hierarchy of multi-soliton solutions for example. The question of correct formulation of integrable boundary problem for 2D Toda lattice (\ref {1.2a}) was considered in \cite{ghz} (see also \cite{hab4}).

Here we are to describe a class of integrable boundary problems for (\ref {1.2a}) and present examples of their explicit solutions. This procedure's ideology is similar to that applied above to KP equation (\ref {1.1}), so we will not get into details.

The Zakharov-Shabat dressing method for (\ref {1.2a}) consists of the following. Introduce operator $\widehat F$ so that

\begin{equation}
\label{5.1}
(\widehat F \Psi)(n)= \sum ^\infty _{k=- \infty} F(n,k)\Psi (k),
\end{equation}

\begin{equation}
\label{5.2}
[\partial _X - A,\ F]=0,\ \ [\partial _Y - L,\ F]=0.
\end{equation}
Suppose that operator $\widehat F$ admits triangular factorization:

\begin{equation}
\label{5.3}
1+\widehat F=(1+\widehat K)^{-1}(1+\widehat K^{-}),
\end{equation}
where

\begin{equation}
\label{5.4}
(\widehat K \Psi)(n)= \sum _{j=n}^{\infty} K(n,j)\Psi (j),\ \ 
(\widehat K^{-} \Psi)(n)= \sum _{j=-\infty}^{n-1} K^-(n,j)\Psi (j).
 \end{equation}
Then specify operators $L^1$ and $A^1$ of difference structure similar to that of operators $L^0$, $A^0$ correspondingly by system

\begin{equation}
\label{5.5}
\begin{array}{c}
(\partial _X -A^1)(1+K)= (1+K) (\partial _X -A^0) \\
(\partial _Y -L^1)(1+K)= (1+K) (\partial _Y -L^0),
\end{array}
\end{equation}
and obtain formula for new solution $u^1$ to 2D Toda lattice via initial solution $u^0$:

\begin{equation}
\label{5.6}
u^1(n)=u^0(n)-\log (1+K(n,n))
\end{equation}
The matrix $K$ solves the difference analogue of Gelfand-Levitan-Marchenko (GLM) equation which arises from (\ref {5.3}), (\ref {5.4}):

\begin{equation}
\label{5.7}
K(n,m)+F(n,m)+\sum _{k=n}^{\infty} K(n,k)F(k,m)=0,\ \ m\ge n.
\end{equation}
To launch the dressing procedure one needs to solve linear equations (\ref {5.2}) which look in the following way:

\begin{equation}
\label{5.8}
\begin{array}{c}
\partial _X F_{nm}=F_{n+1,m}-F_{n,m-1}+F_{nm}(u_X(m)-u_X(n))\\
\partial _Y F_{nm}=w(m)F_{n,m+1}-w(n-1)F_{n-1,m}.
\end{array}
\end{equation}
It is easy to see that system (\ref {5.8}) has solutions $F_{nm}=\Psi (n)\widehat \Psi(m)$ where $\Psi (n)$ solves equations (\ref {5.0}) whereas $\widehat\Psi (n)$ solves the ``conjugate'' system

\begin{equation}
\label{5.9}
\begin{array}{c}
\widehat\Psi _X (n)= u_X (n) \widehat\Psi (n)-\widehat\Psi(n-1),\\
\widehat\Psi _Y (n)= w(n) \widehat\Psi (n+1).
\end{array}
\end{equation}
In this case the desired solutions to GLM equation (\ref {5.7}) take the following form:

\begin{equation}
\label{5.10}
K(n,m)=-\frac{\Psi (n) \widehat \Psi (m)}{1+\sum ^{\infty}_{j=n} \Psi (j)\widehat \Psi (j)}.
\end{equation}
Now one has just to substitute (\ref {5.10}) into formula (\ref {5.6}) to obtain the ``dressed'' solutions, then in virtue of the Definition 1 select solutions which satisfy the integrable boundary constraint.

\section{Integrable boundary problems for 2D Toda lattice.}
\setcounter{equation}{0}

In systems of linear equations (\ref {5.0}) è (\ref {5.9}) we replace variables $(X,\ Y)\rightarrow (x,y)$ via functions $X=X(x,y),\ Y=Y(x,y)$ and rewrite these systems in the following way:

\begin{equation}
\label{6.2}
\begin{array}{c}
\Psi _x (n)= \frac {1}{\Delta}\left\{ f_{22}[-(f_{11}u_x (n)+ f_{12}u_y (n)) \Psi (n)+\Psi(n+1)]\right.\\
\left. +f_{12}w(n-1)\Psi(n-1)\right \} \\
\Psi _y (n)= -\frac {1}{\Delta}\left\{ f_{21}[-(f_{11}u_x (n)+ f_{12}u_y (n)) \Psi (n)+\Psi(n+1)]\right.\\
\left.+f_{11}w(n-1)\Psi(n-1)\right\},
\end{array}
\end{equation}

\begin{equation}
\label{6.3}
\begin{array}{c}
\widehat\Psi _x (n)= \frac {1}{\Delta}\left\{ f_{22}[-(f_{11}u_x (n)+ f_{12}u_y (n)) \widehat\Psi (n)-\widehat\Psi(n-1)]\right.\\
\left. -f_{12}w(n)\widehat\Psi(n+1)\right \} \\
\widehat\Psi _y (n)= -\frac {1}{\Delta}\left\{ f_{21}[-(f_{11}u_x (n)+ f_{12}u_y (n)) \widehat\Psi (n)-\widehat\Psi(n-1)]\right.\\
\left.-f_{11}w(n)\widehat\Psi(n+1)\right\},
\end{array} 
\end{equation}
where $f_{ij}$ are partial derivatives: $f_{11}=x_X,\ f_{12}=y_X,\ f_{21}=x_Y,\ f_{22}=y_Y,\ \Delta =f_{11}f_{22}-f_{21}f_{12}$. Suppose that $u(n)$ solves the integrable boundary problem, therefore due to the Definition 1 the linear correlation

\begin{equation}
\label{6.31}
\widehat\Psi (n)=B(n)\Psi (n)
\end{equation}
must hold on the boundary $x=x_0$ or, in terms of variables $X,\ Y$, on parametric curve $X=X(x_0,y),\ Y=Y(x_0,y)$. Substitute relation (\ref {6.31}) into the second equation of system (\ref {6.3}), using (\ref {6.2}) collect coefficients at functions $\Psi(n),\ \Psi(n\pm 1)$ and obtain explicit form for multiplier $B(n)$:

\begin{equation}
\label{6.4}
\left. B(n)=e^{u(n)}\left( -\frac{f_{21}}{f_{11}}\right)^n d(y)\right |_{x=x_0}
\end{equation}
and for the boundary constraint:

\begin{equation}
\label{6.5}
\left. n\partial _y\left( \log \frac{f_{21}}{f_{11}}\right)+u_y(n)+\frac{2f_{21}}{\Delta}(f_{11}u_x (n)+ f_{12}u_y (n))+\partial _y \log d(y)\right |_{x=x_0}=0,
\end{equation}
where $\partial _y=\partial /\partial y,\ d(y)$ is arbitrary function. We impose also the following constraints:

\begin{equation}
\label{6.6}
f_{11}\ne 0,\ f_{21}\ne 0,\ \Delta \ne 0.
\end{equation}

\section{Explicit solutions to boundary problems.}
\setcounter{equation}{0}

Now we illustrate the method's effectiveness on a few examples.

1. Take the following variable transformation:

\begin{equation}
\label{7.1}
X(x,y)=e^{a(x)+b(y)},\ \ Y(x,y)=e^{c(a(x)-b(y))},
\end{equation}
where $a(x),\ b(y)$ are arbitrary smooth functions, $c$ is non-zero parameter. Then the boundary contour is presented by curve 

\begin{equation}
\label{7.1a}
YX^c=e^{2ca(x_0)}=D=const
\end{equation}
on $(X,\ Y)$-plane. On this contour we have the appropriate boundary constraint which ensues from formula (\ref {6.5}):

\begin{equation}
\label{7.3}
u_x(n)\left| _{x=x_0}\right.=a'(x_0)\left[(1+c)n+\frac{d'(y)}{b'(y)d(y)}\right],
\end{equation}
and the system of linear equations (\ref {6.2}), (\ref {6.3}) reads as follows:

\begin{equation}
\label{7.4}
\begin{array}{c}
\Psi _x (n)= a'(x)X \Psi (n+1)-\frac{1}{2}\left(u_x(n)+\frac{a'(x)u_y(n)}{b'(y)}\right)\Psi(n)\\
-ca'(x)Yu(n-1)\Psi(n-1)\\
\Psi _y (n)= b'(y)X \Psi (n+1)-\frac{b'(y)}{2a'(x)}\left(u_x(n)+\frac{a'(x)u_y(n)}{b'(y)}\right)\Psi(n)\\
+cb'(y)Yu(n-1)\Psi(n-1),
\end{array}
\end{equation}

\begin{equation}
\label{7.5}
\begin{array}{c}
\widehat\Psi _x (n)= ca'(x)Yu(n) \widehat\Psi (n+1)+\frac{1}{2}\left(u_x(n)+\frac{a'(x)u_y(n)}{b'(y)}\right)\widehat\Psi(n)\\
-a'(x)X\widehat\Psi(n-1)\\
\widehat\Psi _y (n)= -cb'(y)Yu(n) \widehat\Psi (n+1)+\frac{b'(y)}{2a'(x)}\left(u_x(n)+\frac{a'(x)u_y(n)}{b'(y)}\right)\widehat\Psi(n)\\
-b'(y)X\widehat\Psi(n-1).
\end{array}
\end{equation}

Now we apply the integration procedure developed above to the boundary problem (\ref {1.2a}), (\ref {7.1a}), (\ref {7.3}) as $d(y)\equiv 0$. Start from the following simple solution to problem (\ref {1.2a}), (\ref {7.1a}), (\ref {7.3}):

\begin{equation}
\label{7.6}
u^0(n)=(1+c)na(x).
\end{equation}
Substitute (\ref {7.6}) into coefficients of systems of linear equations (\ref {7.4}), (\ref {7.5}) and find their solutions:

as $c\neq 1$ 

\begin{equation}
\label{7.7}
\begin{array}{c}
\Psi (n)=\exp \left( -\frac{n}{2}(1+c)(a(x)+b(y))\right)\\
\times\sum _{j=1}^N k_jp^n_j \exp \left[ \frac{2e^{\frac{1-c}{2}b(y)}}{1-c}\left( p_j e^{\frac{1-c}{2}a(x)}+\frac{c}{p_j}e^{\frac{c-1}{2}a(x)}\right)\right]
\end{array}
\end{equation}

\begin{equation}
\label{7.8}
\begin{array}{c}
\widehat\Psi (n)=\exp \left( \frac{n}{2}(1+c)(a(x)+b(y))\right)\\
\times\sum _{j=1}^N l_jq^n_j \exp \left[ -\frac{2e^{\frac{1-c}{2}b(y)}}{1-c}\left( \frac{1}{q_j} e^{\frac{1-c}{2}a(x)}+cq_j e^{\frac{c-1}{2}a(x)}\right)\right]
\end{array}
\end{equation}
where $k_j,\ l_j,\ p_j,\ q_j,\ j=1,2,...,N$ are arbitrary parameters. Now impose the integrability constraint

\begin{equation}
\label{7.9}
\widehat\Psi (n)=B(n)\Psi (n)|_{x=x_0},
\end{equation}
where

\begin{equation}
\label{7.10}
B(n)=(-1)^ne^{n(2a(x_0)-\log c +(1+c)b(y))},
\end{equation}
and obtain relations on parameters:

\begin{equation}
\label{7.11}
l_j=k_j,\ q_j=-\frac{p_j}{c}e^{(1-c)a(x_0)}.
\end{equation}
Then assume solution to GLM equation in the form (\ref {5.10}):

\begin{equation}
\label{7.12}
K(n,m)=-\frac{\Psi (n)\widehat \Psi (m)}{1+\sum _{j=n} ^{\infty}\Psi (j)\widehat \Psi (j)},
\end{equation}
and substitute it into the formula

\begin{equation}
\label{7.13}
u(n)=u^0(n)-\log (1+K(n,n)),
\end{equation}
which yields the ``dressed'' solution to 2D Toda lattice that may be formally associated with $N$-soliton solutions. In the simplest $(N=1)$ case we have

$$
K(n,n)=-\left\{\nu ^{-n}\exp \left [\frac{2e^{\frac{1-c}{2}b(y)}}{c-1}\left( \frac{c}{p}(1-\nu)e^{\frac{c-1}{2}a(x)}\right.\right.\right.
$$
$$
\left.\left.\left.+p\left(1-\frac{1}{\nu}\right)e^{\frac{1-c}{2}a(x)}\right)\right]+\frac{1}{1-\nu}\right\}^{-1},
$$
which after substitution into (\ref {7.13}) gives the simplest ``dressed'' solution to the boundary problem. Now present the key formulae in terms of initial variables $X,\ Y$.

Boundary constraint:

\begin{equation}
\label{7.14}
Xu_X+ñ Y u_Y =(1+c)n
\end{equation}

boundary contour

\begin{equation}
\label{7.15}
X^c Y=e^{2c a(x_0)}=D=const,
\end{equation}

the simplest ``dressed'' solution to 2D Toda lattice is shown by formula (\ref {7.13}), where

\begin{equation}
\label{7.16}
u^0(n)=\frac{1}{2}(1+c)n\left( \log X +\frac{1}{c}\log Y\right),
\end{equation}

$$K(n,n)=$$
\begin{equation}
\label{7.17}
\left[\frac{1}{\nu -1}-\nu ^{-n}\exp \left(\frac{2}{c-1}\left(\frac{c}{p} (1-\nu)Y^{\frac{c-1}{2c}}+p\left(1-\frac{1}{\nu}\right)X^{\frac{1-c}{2}}\right)\right) \right]^{-1},
\end{equation}
$$\nu =- \frac{p^2}{c}D^{\frac{1-c}{2c}}.$$

In the case $c=1$ appropriate formulae take the following form:

boundary constraint

$$Xu_X+Yu_Y=2n$$

boundary contour

$$XY=D,$$
the simplest ``dressed'' solution to 2D Toda lattice is shown by formula (\ref {7.13}), where 

\begin{equation}
\label{7.20}
u^0(n)=n(\log X+\log Y),
\end{equation}

\begin{equation}
\label{7.21}
K(n,n)=-\left[(-p^2)^{-n}\left(\frac{Y}{X}\right)^{p+\frac{1}{p}}+\frac{1}{1+p^2}\right]^{-1}.
\end{equation}

2. Now examine the variable transformation

\begin{equation}
\label{7.22}
X=e^{a(x)}\sin b(y),\ \ Y=ce^{a(x)}\cos b(y),
\end{equation}
which forms closed contour. Calculations which determine corresponding boundary problem and its explicit solutions are similar to those mentioned above, so we present here just final outcome in the initial variables:

boundary constraint

$$
Yu_X+c^2Xu_Y+n\left(\frac{Y}{X}+c^2\frac{X}{Y}\right)=0,
$$

boundary contour

$$
X^2+\frac{Y^2}{c^2}=e^{2a(x_0)}=D,
$$

the ``dressed'' solution is presented by formulae

$$
u^0(n)=-n\log (XY),
$$

$$K(n,n)=$$

\begin{equation}
\label{7.29}
-\left\{\left(-\frac{p^2}{c^2}\right)^{-n}\exp \left[\frac{1}{2}\left(p+\frac{c^2}{p}\right)\left(\frac{Y^2}{c^2}-X^2\right)\right]+\frac{1}{1+\frac{p^2}{c^2}}\right\}^{-1},
\end{equation}
$p$ is arbitrary parameter.

3. At last we have the example of regular solution. Put $c=-1$ in the formulae (\ref {7.1})-(\ref {7.17}) and obtain:

boundary constraint

$$
Xu_X-Yu_Y=0,
$$
boundary contour

\begin{equation}
\label{7.29a}
Y/X=e^{2a(x_0)}=D.
\end{equation}

The initial solution is chosen as a constant: $u^0(n)=u^0$ and

\begin{equation}
\label{7.30}
K(n,n)=\left[\frac{1}{\nu -1}-\nu^{-n}\exp \left(\frac{1-\nu}{p}Y-p\left(1-\frac{1}{\nu}\right)X\right)\right]^{-1},
\end{equation}
where $\nu=\frac{p^2}{D}$. The formula (\ref {7.13}) yields solution regular as $0<\nu<1$. This solutions presents an ordinary soliton that travels along the straight boundary line (\ref {7.29a}), so no reflection effects occur.

\section{Conclusion.}

In this paper we show that:

1) integrable in force of the Definition 1 boundary problems for Kadomtsev-Petviashvili equation and 2D Toda lattice can be solved explicitly. We develop the method and show its effectiveness on concrete examples

2) the set of integrable boundary problems that consist of boundary countours and appropriate boundary constraints is large enough. This fact is illustrated by formulae (\ref {3.5}), (\ref {3.6}), (\ref {6.5}), (\ref {6.6}).

The author thanks Professor I.T. Habibullin for valuable consultations. The investigation was partially supported by grants of the Program of RAS Presidium  ``Fundamental problems of nonlinear dynamics'' and  Russian Foundation for Basic Research 10-01-00088-a.

\pagebreak


\begin {thebibliography} {99}

\bibitem{kadpet} B.B. Kadomtsev and V.I. Petviashvili// Soviet Math. Dokl., v. 192, n. 4, p. 753-75 (1970)

\bibitem {mik1} Mikhailov A.V.// JETP Letters (1979), v.39, n.7, P. 443-448

\bibitem {teorsol} S.P. Novikov, S.V. Manakov, L.P.Pitaevski and V.E. Zakharov. Soliton theory. Nauka, Moscow 1980

\bibitem {dr} V.S. Dryuma// JETP Letters (1974), v. 19, n. 12, n. 753-755

\bibitem {zash} V.E. Zakharov and A.B. Shabat// Funkz. analiz i ego pril. (1974), v. 8, n. 3, p. 43-53

\bibitem {skl} E.K. Sklyanin// Funkz. analiz i ego pril. 21:2 (1987), p. 86-87

\bibitem {hab1} I.T. Habibullin// Theor. Math. Physics, 91:3 (1992),  363-376

\bibitem {hab2} I.T. Habibullin and E.V. Gudkova// Funkz. analiz i ego pril. 38:2 (2004),  71-83

\bibitem {hab3} I.T. Habibullin and E.V. Gudkova// Theor. Math. Physics, 140:2 (2004),  230-240

\bibitem {aggh} V. Adler, B. Gurel, M. Gurses and I.T. Habibullin // J. Phys. A: Math. Gen. 30 (1997), p. 3505-3513

\bibitem {mik2} Mikhailov A.V.// Physica D (1981), V.3, n.1,2, P.73-117

\bibitem {ueta} K. Ueno and K. Takasaki, Toda Lattice Hierarchy, Adv. Stu. in Pure Math. 4 (1984) 1

\bibitem {naka} Nakamura A. // J. of the Phys. Soc. Japan, (1983), V.52, n.2, P.380-387

\bibitem {bms} M.V. Babich, V.B. Matveev and M.A. Sall// Zapiski nauch. seminar. LOMI (1985), v.145, p. 34-45

\bibitem {ghz} Gurses M., Habibullin I.T. and Zheltukhin K.// J. of Math. Physics (2007), V.48, 102702

\bibitem {hab4} Habibullin I.T. and A.N.Vil'danov, Integrable Boundary Conditions for Nonlinear Lattices, CRM,  Proceedings and Lecture Notes, Amer. Math. Soc., Providence, v. 25, 2000, p. 173 - 180.

\end {thebibliography}
\end {document}